\documentclass{elsart}
\usepackage{graphicx}
\usepackage{amssymb}
\usepackage{epsfig}

\newcommand{\bea}{\begin{eqnarray}}
\newcommand{\eea}{\end{eqnarray}}
\newcommand{\be}{\begin{equation}}
\newcommand{\ee}{\end{equation}}
\newcommand{\nn}{\nonumber}
\newcommand{\btof}{\gamma_{b \rightarrow f}}
\newcommand{\ftob}{\gamma_{f \rightarrow b}}

\newcommand{\ave}[1]{\mbox{$<\!\!#1\!>$}}

\begin{document}
\begin{frontmatter}

\title{Collective Effects in Models for Interacting Molecular Motors 
and Motor-Microtubule Mixtures}
\author{Gautam I. Menon}
\ead{menon@imsc.res.in}
\address{The Institute of Mathematical Sciences, C.I.T. Campus,\\ Taramani,
         Chennai 600 113, India}
\begin{abstract}
Three problems in the statistical mechanics of models
for an assembly of molecular motors interacting
with cytoskeletal filaments are reviewed.  First,
a description of the hydrodynamical behaviour of
density-density correlations in fluctuating ratchet
models for interacting molecular motors is outlined.
Numerical evidence indicates that the
scaling properties of dynamical behavior in such
models belong to the KPZ universality class.  Second,
the generalization of such models to include boundary
injection and removal of motors is provided. In common
with known results for the asymmetric exclusion
processes, simulations indicate that such models
exhibit sharp boundary driven phase
transitions in the thermodynamic limit.  In the third
part of this paper, recent progress towards a continuum
description of pattern formation in mixtures of motors
and microtubules is described, and a non-equilibrium
``phase-diagram'' for such  systems discussed. 
\end{abstract}
\end{frontmatter}
\section{Introduction}\label{sec:intro}
Living systems exhibit a remarkable variety of
non-equilibrium steady states. Problems associated
with the modelling of such states include the
description of the non-equilibrium behaviour of
membranes driven by active pumps\cite{madan},
hydrodynamic approaches to the motion of
self-propelled objects\cite{sriram0,sriram}, the
theory of pattern formation in a variety of biological
contexts\cite{meinhardt} and models for intracellular
transport processes associated with the motion of
molecular motors\cite{debashish}. This last problem
has attracted the attention of statistical physicists
in recent years, since the simplest models for such
systems have several advantages: they are exactly
solvable even in the presence of interactions, easy
to generalize, relatively straightforward to simulate
and closely related to models studied extensively in
the context of traffic flow\cite{debashish}.

It is useful to develop intuition with simple models,
requiring only that it should be possible to add the
requisite biological detail incrementally. This would
then allow progressively more accurate descriptions
to be constructed within a sequence of increasingly
refined models. This paper reviews some calculations
which explore the middle ground between extremely
simplified statistical mechanics models for the
motion of interacting molecular motor proteins  --
the asymmetric exclusion process and generalizations
-- and marginally more realistic models for the motion
of individual motor proteins generalized to accomodate
motor-motor interactions\cite{yashar}. It also reviews
some recent work on the hydrodynamic description of
pattern formation in mixtures of molecular motors and
microtubules\cite{sumithra}.

Molecular motors are a class of biological machines
which function within cells\cite{mehta}.  Such motors,
proteins such as kinesins, myosins and dyneins,
move unidirectionally on one-dimensional ``tracks''
while hydrolysing adenosine tri-phosphate (ATP).
These tracks are components of the cytoskeleton in
eukaryotic cells, an extended dynamic network formed
through the polymerization and crosslinking of tubulin
and actin monomers to form microtubules and actin
filaments\cite{howard}.  The cytoskeleton helps the
cell anchor to substrates, to move and to divide,
and lends it mechanical and structural rigidity.
In addition, this network defines paths for molecular
motors to transport cargo to different parts of
the cell\cite{howard}.

The asymmetric nature of motor motion along the
cytoskeleton derives from the asymmetry of the
constituent monomeric units of the track.  Microtubules
(equivalently, actin filaments) can be idealised as
periodic, one-dimensional, rigid structures, this
periodicity following from their polymeric nature.
The violation of detailed balance necessary in order
for the motor to exhibit directed motion comes from the
transduction of the chemical energy obtained from ATP
hydrolysis into mechanical work\cite{prostRMP}.  The
stochastic uptake of ATP is one source of random noise
in the problem; the other is the thermal noise which
dominates all biological systems at cellular scales.
Molecular motors thus act as {\it Brownian rectifiers}
in exhibiting a non-zero drift velocity in the absence
of a net time-averaged force.  The problem of modelling
molecular motors can therefore be placed in the more general
context of ``Brownian motor'' or ``thermal ratchet''
models for the extraction of useful work from thermal
fluctuations\cite{prostRMP,reimann}.

We consider the ``fluctuating potential'' or ``flashing
ratchet'' model for Brownian motors\cite{prostRMP}. In
this model, a single motor, idealized as a point
object moving in one dimension, is driven by stochastic
forces uncorrelated in space and time and drawn from a
Gaussian distribution with zero mean.  The motor also
experiences a force derived from the gradient of a
time-dependent potential. The time dependence of this
potential is generated by switching randomly between
two states, one in which the potential is an asymmetric
sawtooth -- say with the longer leg of the potential
to the left -- and the other in which it is flat.

\begin{figure}
\includegraphics[width=\columnwidth]{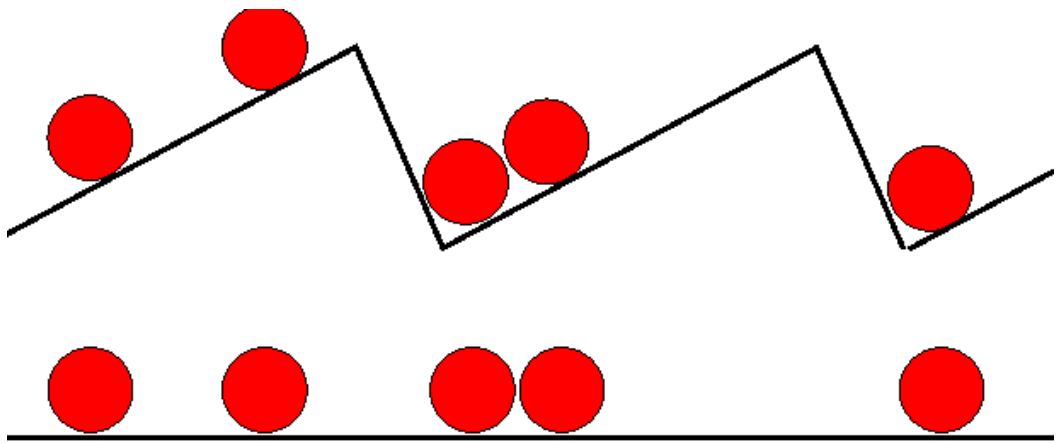}
\caption{The continuum potential corresponding to the
two states (non-trivial and flat) of the potential,
showing the locations of several motors interacting
through a hard-core repulsive potential.
}
\end{figure}

In the ``off' or ``flat'' state of the potential,
particles diffuse isotropically.  When the potential
is switched to the ``on'' state, a particle is more
likely to be found in a region of space where it
experiences a net force to the left than to the right,
given that the potential lacks right-left symmetry.
Repeated cycling between ``off'' and ``on'' states
generates net motion. The generation of directed motion
occurs through a subtle mechanism: in any one of the
potential states, given sufficient time to equilibrate,
no net current can flow provided the microscopic jump
rates obey detailed balance. The breaking of detailed
balance overall arises from the non-equilibrium, 
time-dependent {\em switching} between potential states 
and not from the choice of hopping rates in any one of 
these states \cite{prostRMP}.

In modelling biological motors, the asymmetric
potential encodes the energy associated with an
internal state of the motor as a function of its
position along the track\cite{prostRMP}. This state
represents a particular conformation of the motor
protein. ATP consumption induces transitions between
states. A motor molecule thus has two degrees of
freedom, a spatial coordinate and an internal (state)
coordinate.  A faithful representation of the internal
states of the motor and track thus involves specifying
a large number of continuous periodic potentials
representing the energy of a molecular motor at
location $x$ as a function of its state $s$.

The simplified model described here retains only two
states of the motor-track complex. Such ``minimal''
models may be expected to capture some of the relevant
complexity of the real biological system, at least
in the limited contexts in which we will study them. It is
straightforward to generalize the models described
here.  In particular, one could allow motors both to
detach from the filament and diffuse within the ambient
solvent as well as to reattach with prescribed rates.
One could also allow for motors of finite extent,
by extending the exclusion constraint to sites which
neighbour the one occupied by the motor.

The effects of cooperativity between motors has
been a source of rich physics in recent years
\cite{prostRMP,frank,frank1,camalet,jayan}.  Such
effects, in biological models for the action of the
myosin motors involved in muscle contraction, have been
incorporated by coupling motors through intervening
elastic elements. Thus, in addition to forces derived
from the external potential and Gaussian noise, a
motor feels a force due to its elastic interactions
with its neighbours. A mean-field analysis of the
effects of such elastic couplings yields a variety
of novel phenomena, such as Hopf bifurcations and
spontaneous oscillations of the current\cite{prostRMP}.
However, another class of interaction effects which
operate between elastically decoupled motors can
be envisaged: the steric hindrance (or, in general,
any short-range interaction) experienced by motors
translocating on a filament when they approach each
other\cite{vicsek,yashar,nishinariprl}.

How can the fluctuating potential model be generalized
to incorporate the effects of interactions between many
motors moving on a cytoskeletal filament?
The simplest such interaction is a hard-core
interaction between two motors, as illustrated
schematically in Fig 1.  On the lattice, this
constraint is simply implemented by allowing only one
particle to occupy a given lattice site at a time.
We define simple lattice models incorporating such
interactions and study these models with boundary
conditions (periodic) which conserve the total number
of motors in Section II.  We also discuss a simple
mean-field theory and its prediction for currents
and density profiles within a single period of
the potential.  A second set of results, presented
in Section  III, relate to the existence of phase
transitions in such systems induced by the effects of
adding (subtracting) motors at the boundaries of the
(open) one-dimensional chain at prescribed rates\cite{krug}.  
Our numerically calculated phase diagram closely resembles
the phase diagram of the partially asymmetric exclusion
process with boundary driving\cite{derrida,robin}.

The fact that these two systems should be related
at the level of hydrodynamic correlations was
conjectured in Ref.\cite{yashar} and 
used to predict that density-density correlations
in the steady state of the interacting motor system
should obey scaling with exponents belonging to the
KPZ universality class.  The fact that these systems
show the same type of phase transitions as a function
of boundary conditions is further evidence of the
close relationship between these models, despite
the far greater complexity (multiplicative time and
space-dependent noise at the level of the microscopic
hopping rates) in the Brownian motor model.

A third set of results reviewed in this paper
(Section IV) relate to the modelling of the
patterns which form when motor complexes are
mixed with microtubules and supplied with ATP, in
a quasi-two-dimensional geometry\cite{sumithra}.
These patterns include structures such as asters,
vortices and aster-vortex mixtures as well as
disordered states. Understanding the generic features
of these states, the sequence of transformations
between them as a function of the motor density and
the interactions which contribute to the formation
of such self-organized structures, is believed
to be a crucial part of understanding the physics
behind the formation of a cellular-scale pattern
universal to all eukaryotic cells, the mitotic spindle
\cite{nedelecsurreymaggsleibler,surreynedelecleiblerkarsenti,nedelecsurrey,nedelecsurreymaggs,nedelecjcb,nedelecsurreykarsenti,heald}.
The concluding section, Section V, discusses some
general features of the results, suggesting that
the general attribute of ``physical robustness'',
a robustness of the non-equilibrium steady states
obtained in such models towards a wide class of
physically relevant perturbations, may be biologically
relevant.

\section{KPZ Scaling in Interacting Ratchet Models}\label{sec:model}
In the simplest continuum
versions of the fluctuating potential models,
individual motor particles see a time-dependent
potential $V(x,t)=\eta(t)U(x)$, in addition to random
Brownian forces with zero mean value. Here $U(x)$
is periodic with period $\ell$ {\it i.e.} $U(x+\ell)
= U(x)$ and an asymmetric function of $x$ {\it i.e.}
$U(x)\neq U(-x)$.  The time dependence of $V(x,t)$ is
governed by a (stochastic or deterministic) switching
function $\eta(t)$ which takes the values 0 and 1. We
assume $U(x)$ to be of the sawtooth form
\begin{eqnarray}
U(x) &=& ax ~~~~~~~~~~~~~(0 \leq x \leq \omega \ell), \nonumber \\
     &=& b(\ell-x)~~~~~~(\omega \ell \leq x \leq \ell)
\label{U(x)}
\end{eqnarray}
with $a,\,b>0$, $a\omega \ell = b \ell(1 -\omega)$ and $\omega < 1$.
The switching of the potential occurs independently of the state of the
motors, thus breaking detailed balance.  Together with the lack of
reflection symmetry in $U(x)$, this switching generates a net particle
current.  

\begin{figure}
\includegraphics[width=\columnwidth]{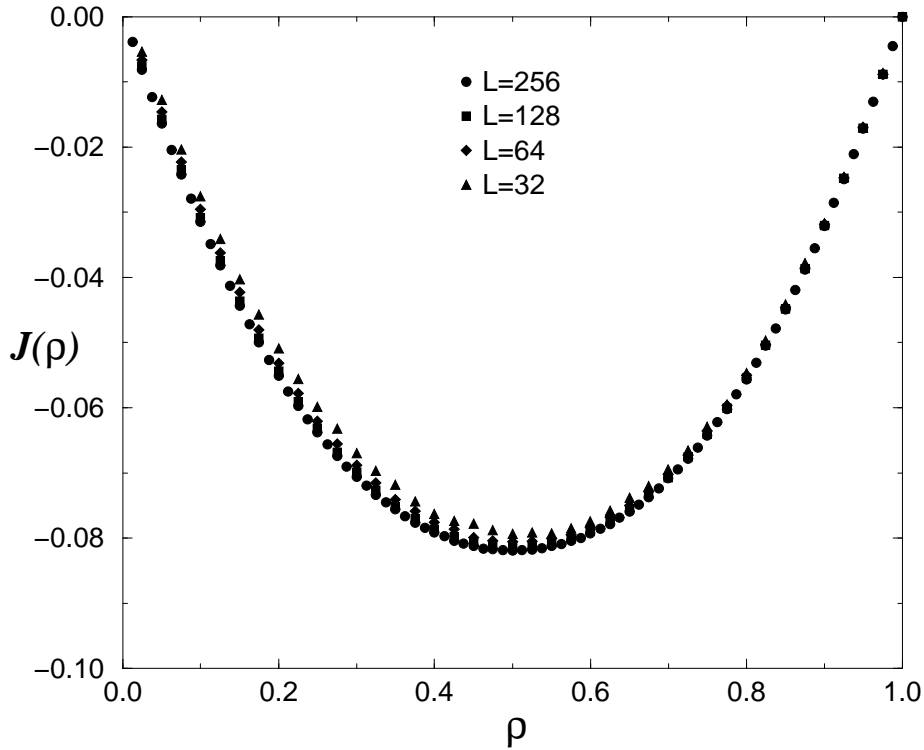}
\caption{Dependence of the steady-state current on the 
density $\rho$ of particles for $32\leq L\leq 256$. 
}
\label{fig:current1}
\end{figure}

\begin{figure}
\includegraphics[width=\columnwidth]{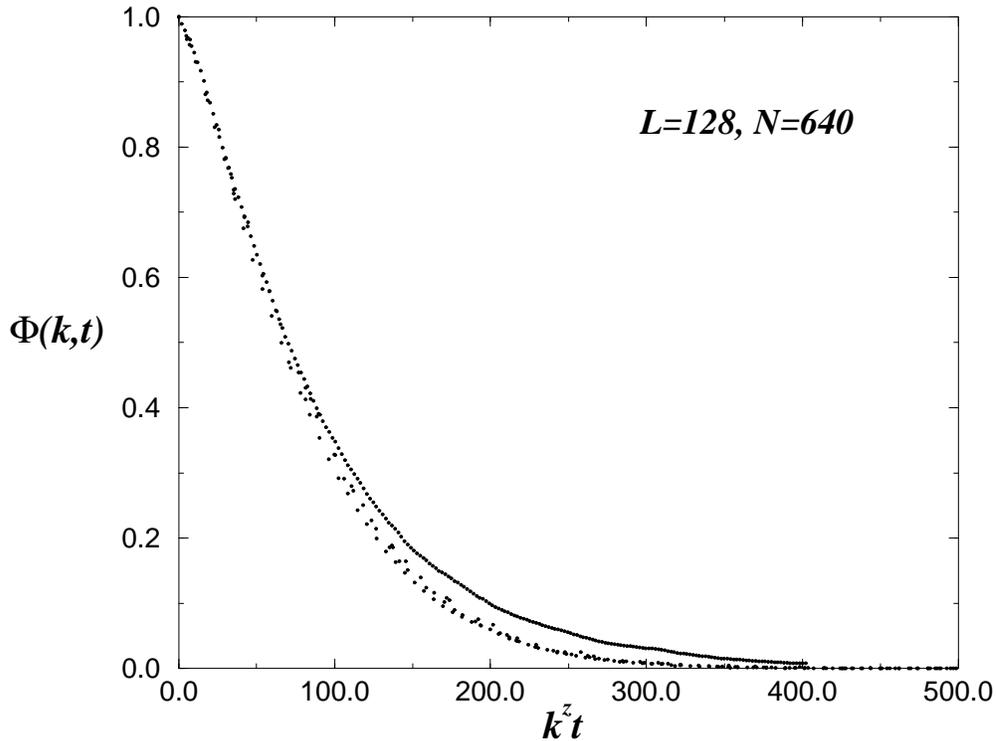}
\caption{The relaxation function $\Phi(k,t)$ 
for $L=128$ and $\rho=0.5$ plotted as a function of the 
scaled variable $k^zt$ for $z=1.60$ for the five smallest 
values of $k=2\pi j/L$, with $j=1,2,3\ldots$. The lowest
k-value splits off whereas data for all higher $k$
fall on the same branch. }
\label{fig:phi1}
\end{figure}

\begin{figure}
\includegraphics[width=\columnwidth]{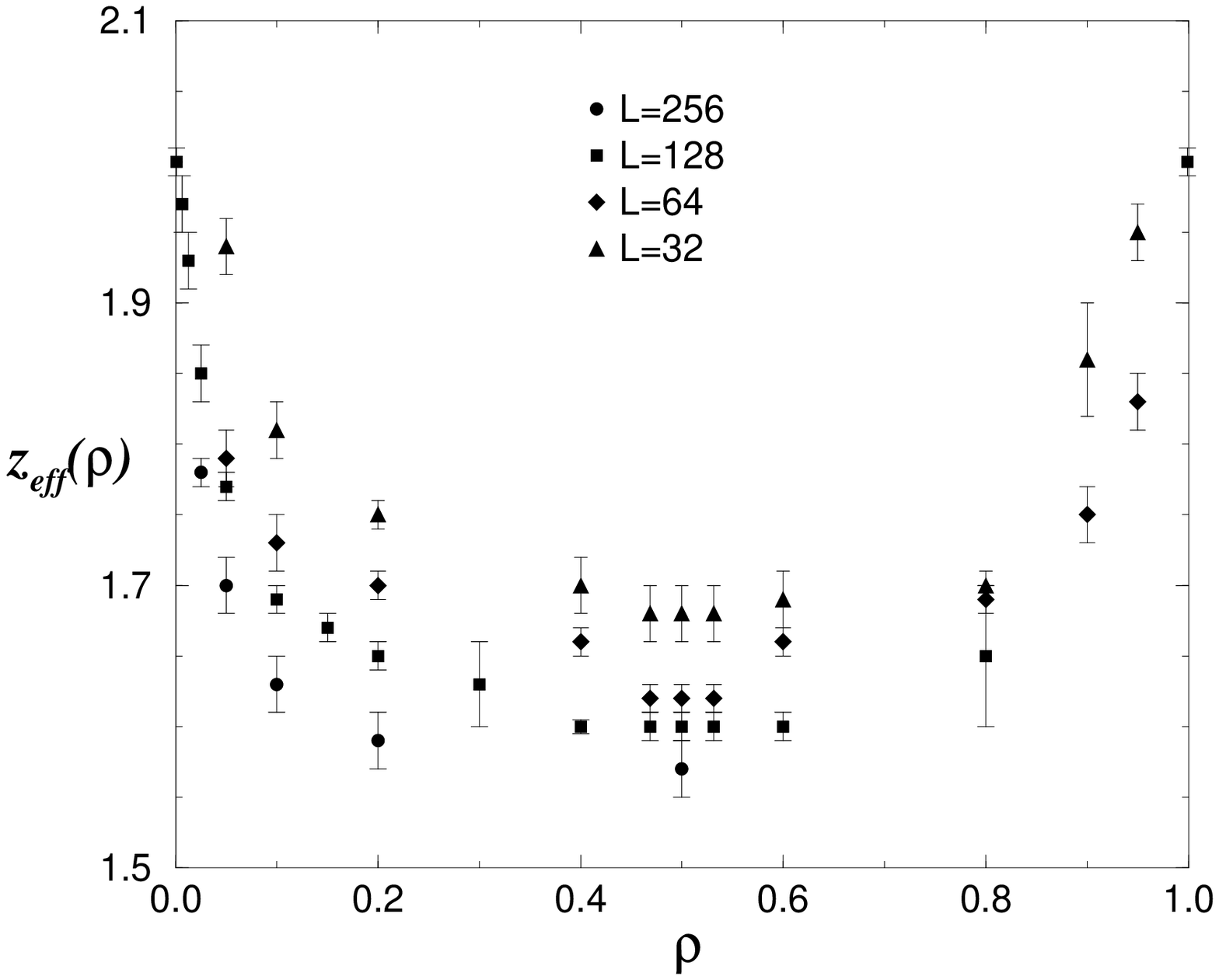}
\caption{The effective exponent $z_{eff}(N,\rho)$ plotted as function of $\rho$
for systems of size $32\leq L\leq 256$ Note the systematic decrease
of $z_{eff}$ as $L$ is increased at constant density. 
}
\label{fig:zeff1}
\end{figure}

A useful simplification is the discretization of
the periodic potential in space to convert the
continuum problem into a lattice one. (This is not
a unreasonable simplification, since the biological
motor appears to undergo a sequence of {\em discrete}
conformational changes, each of which is coupled to a
partial translocation across the period.)  Each period
of the potential is divided into $W$ lattice sites,
all of which are assigned to the segment of the
potential with positive slope.
The length of the system, $L$, is measured in periods
of the sawtooth.  The maximum height of the sawtooth
potential is $V$ and we define $r=\exp(-\frac{V}{k_B
T (W-1)})$.  Finally, the parameters $P_{01}$ and
$P_{10}$ represent the probabilities that the potential
changes from $\eta = 0$ to $\eta = 1$ and vice versa.

The transition rates between the configurations of motors
are chosen to satisfy
detailed balance with Metropolis rates,
$P\left(\{\bar{\sigma'}\}\to\{\bar{\sigma}\}\right) = \min(1,
\exp[H(\{\bar{\sigma'}\})-H(\{\bar{\sigma}\})])$. Here $\{\sigma\}$
indexes allowed configurations of the motors on the lattice and
$H(\{\sigma)\}$ is the energy of the configuration.  
In terms of these parameters the hopping probabilities for motor
particles on the sawtooth are 
\begin{eqnarray}
\label{eq:hops}
P(i \rightarrow i+1) & = & \left\{ \begin{array}{ll}
					\frac{r}{1+r} & \mbox{if $0 \leq i < W-1$} \\
					\frac{1}{1+r^{W-1}} & \mbox{if $i = W-1$,} 
				   \end{array}
				   \right. \nonumber \\
P(i+1 \rightarrow i) & = & \left\{ \begin{array}{ll}
					\frac{1}{1+r} & \mbox{if $0 \leq i < W-1$} \\
					\frac{r^{W-1}}{1+r^{W-1}} & \mbox{if $i = W-1$,}
				   \end{array}
				   \right.
\end{eqnarray}
where we have indexed the lattice sites from $0$ at the potential minimum.
The hopping probabilities for left(right) jumps on the flat potential 
equal 1/2 at every site.

To generalize models for single Brownian motors
to a finite number $N_p$ of interacting motors on
a one-dimensional lattice, we assume that these
motors hop to unoccupied nearest neighbour sites
with rates determined by the discretized potential
in Eq. (\ref{eq:hops}), exactly as they would in the
non-interacting case.  The only interaction between
these motors is thus a hard-core repulsion which
prevents them from occupying the same lattice site. Our
discretization and the hard-core constraint ensures an
upper bound on the number of motors which can occupy
the lattice.

We consider periodic boundary conditions in the
calculations described in this section. Our motors thus
move on a ring.  An elementary step consists of either
an attempt of a particle to hop to a neighbouring
site or an attempt to switch the potential state
globally. The results of Ref. \cite{yashar} were
obtained by a procedure which involved changing the
potential state globally. Flipping this state locally
does not alter the conclusions qualitatively and the
quantitative changes are small.

Numerical results for the model defined above (Model I
of Ref. \cite{yashar}) are obtained in the following way:  
We use $W=10$ lattice sites per period
of the asymmetric sawtooth in all our simulations,
varying the system size $L$ = 24,48,64,128, 256 and 512.
We took $r=0.05$, $P_{01} = 0.03$ and
$P_{10} = 0.04$, where $P_{01}$ and $P_{10}$ are the
probabilities that the potential state goes from flat
to non-flat and vice versa.  Typically, the system
equilibrates over $5 \times 10^4$ MCS before data for
currents and correlation functions are recorded. These
quantities are averaged over $5\times 10^3 - 10^4$
configurations.

Fig. 2 illustrates the fundamental relation (the
functional dependence of the current on the density) of
the model.  We examine principally the scaling properties
of the intermediate scattering function $S_{\rho}(k,t)$ 
defined by
\begin{equation}
S_{\rho}(k,t) = \frac{1}{N}<\delta\rho(k,0)\delta\rho(-k,t)>,
\end{equation}
where $N$ is the number of particles, $k$ is $2n\pi/L,
n = 0,\pm1,\pm2,\ldots,L/2$, and $L$ is the system size.
Here $\delta\rho(k,t)$ is the Fourier component with wave
vector $k$ of the deviation from the mean local density, 
$\delta\rho(x,t) = \rho(x,t) 
- <\rho(x)>$, with the brackets $<\cdot>$ denoting a time
average. 

Transforming the density field $\rho(x,t)$ to a 
``height'' field $h(x,t)$ via $\rho(x,t) = 
\partial_x h(x,t)$, and imposing helical boundary conditions
on $h(x,t)$ to satisfy the constraint 
$\int_0^L \rho(x,t) = h(L) - h(0) = N$, $S_\rho(k,t)$ can 
be related to the structure factor
$S(k,t) = \langle\delta h(k,0)\delta h(-k,t) \rangle$
where $\delta h(k,t)$ is the Fourier transform of
$h(x,t) - \langle h(x) \rangle $. For small $k$ and large $t$, 
{\it i.e.} in the hydrodynamic limit, if 
$S(k,t)$ exhibits dynamical scaling, we can write
$S(k,t) \sim k^{-2 + \eta}F(k^zt)$,
where $\eta$ and $z$ are scaling exponents and $F$ 
is a scaling function.

In Ref.\cite{yashar}, it was conjectured that the
dynamical scaling properties of the ``height'' field 
should be described 
by the Kardar-Parisi-Zhang \cite{KPZ} equation, 
a non-linear Langevin equation of the form,
\begin{equation}
\label{eq:KPZ}
\frac{\partial h}{\partial t}
 = \nu \nabla^{2}h + \frac{\lambda}{2}(\nabla h)^{2} + \zeta({\bf x},t),
\end{equation}
which is known to describe the long-time,
long-wavelength behavior of a number of 
nonequilibrium systems~\cite{barabas}.
This equation is written in a form appropriate for surface models where
$h({\bf x},t)$ is the height, relative to a $d$-dimensional
substrate, of a growing interface and $\zeta({\bf x},t)$ represents
white noise.  
For the KPZ equation, owing to the existence of
a fluctuation-dissipation theorem, the exponents
can be obtained exactly for $d=1$ and take on the
values $z=3/2$ and $\eta=0$.
The connection is made in the following way:
Coarse-grain microscopic
configurations of such models in space and time. At
spatial scales larger than the repeat distance $\ell$
of the periodic potential and for time scales much
larger than the typical time-scale $\tau$ over which
the potential changes, the system will appear to
have a constant density on average, as well as a
constant current.  Superimposed on this constant
density are spontaneous fluctuations which obey a
local conservation law. The effects of interparticle
interactions at the largest length scales can be
summarized in the following observation: These
density fluctuations are convected with a speed,
the ``kinematic wave speed'', which depends on their
magnitude.

Consider now the statistical mechanics of an unrelated
model, that of the stochastic dynamics of a finite
density $\rho$ of hard-core particles on a line,
which hop individually with rate $(1+\epsilon)/2$ to
the right and $(1-\epsilon)/2$ to the left, provided
the excluded volume constraint is satisfied is known
as the ``asymmetric exclusion process'' (ASEP) for
$\epsilon \neq 0$. The ASEP  has a net particle
current $J = \epsilon\rho(1-\rho)$. The symmetry
breaking which results in a constant current in the
ASEP is an explicit consequence of the asymmetry in
the hopping rates.  This symmetry breaking is to be
contrasted to the more subtle symmetry breaking in
the case of the ratchet models.

Density-density correlations for the
ASEP are known to be governed by KPZ
exponents\cite{beijeren,dhar,gwaspohn}.  Since both
models share the feature expected to be most relevant
to a hydrodynamic description -- the existence of
a non-trivial density-dependent current -- it is
reasonable to conjecture that they should belong to
the same universality class, irrespective of the fact
that the detailed origin of the symmetry breaking is
different in each case.

In our simulations we measured the relaxation 
function~\cite{beijeren}
\begin{equation}
\label{eq:Phi}
\Phi(k,t) \equiv \frac{\ave{\hat{\rho}(-k,0)\hat{\rho}(k,t)}-
\ave{\hat{\rho}(-k,0)}\ave{\hat{\rho}(k,t)}}
                 {\ave{\hat{\rho}(-k,0)\hat{\rho}(k,0)}-\ave{\hat{\rho}(-
k,0)}\ave{\hat{\rho}(k,0)}}=\frac{S_{\rho}(k,t)}{S_{\rho}(k,0)},
\end{equation}
which is just the Fourier transform of
$\sum_i (\rho_i(t)-\rho_i(0))(\rho_{j+i}(t)-\rho_{j+i}(0))$,
normalised by its value at $t=0$, an arbitrarily chosen time in the 
steady state.
The data for $\Phi(k,t)$ for a given density and
system size were plotted as a function of the scaled
variable $k^zt$ for various $z$ and the value of $z$
that provided the best collapse of the data by visual
inspection was taken to be $z_{eff}$. Examples of this
data collapse are shown in Fig \ref{fig:phi1}
at half filling for $L=128$ for $z_{eff}=1.60$. 
The relaxation function generically has
two distinct branches: For the smallest value of $k$,
$k=2\pi/L$, the relaxation function decays more slowly
than for larger values of $k$.  For $j>1$ the
data collapse to quite high accuracy onto a single
curve. This separation of the relaxation function into
two branches (with $j=1$ the special case) is also a
feature of the single step model\cite{plischke2}.

We have carried out this analysis systematically for
$L=32$ to $L=256$ as function of $\rho$. The effective
exponents $z_{eff}(N,\rho)$ are
plotted as functions of $\rho$ in Fig. \ref{fig:zeff1}.
There is a systematic decrease of the effective exponent $z_{eff}$
as function of increasing $L$ for all $\rho\neq 0$,
with the smallest values occuring at $\rho=0.5$, as
expected on theoretical grounds.  The smallest value of
the critical exponent, $z$, from this data was found to
be $1.58 \pm .01$ for $L=256$.  Extrapolating to the
thermodynamic limit, $z(\rho)$ approaches the function:
\begin{eqnarray*} z & = & \left\{ \begin{array}{ll}
		 \frac{3}{2} & \mbox{\ \ for $0 <
		 \rho < 1$} \\ 2  & \mbox{\ \ for
		 $\rho = 0,1$}.  \end{array}
	\right.\,, \end{eqnarray*}
consistent with the predictions of the KPZ equation.

\subsection{Mean-Field Approximation}
\label{sec:mft}
Assuming periodic boundary conditions, all periods
of the potential are equivalent in steady state. By
translational invariance, it is then sufficient
to solve for the steady state density fields and
steady-state currents within a single period.
If a site is to be updated, which is the case with
probability $1/2$ in each elementary time step, the
density $\rho_i(t)$ at site $i$ at time $t$ changes
through hops on and off that site from neighbouring
sites.  These hops can only occur if the hard-core
constraint is satisfied.  Allowed hops occur with the
(microscopic) probabilities $p^L_i(t)$ and $p^R_i(t)$
for hops to the left ($p^L_i(t))$ or right $(p^R_i(t))$
at site $i$ at time $t$. Given our algorithm, in which
an attempt is made to update either a particle or
a potential state at each time step and never both,
the assignment of particle hopping rates at time $t$
is unambiguous.  However, how these rates are to
be interpreted in the time-continuum limit is not
straightforward and will be dealt with explicitly in
what follows.

The update processes for a site $i$ at time $t$ can be written purely 
in terms of the local density 
fields in the neighbourhood of that site. The density 
field $\rho_i(t+1)$ at time $t+1$ is given by :
\begin{eqnarray}
&& \left\{ \begin{array}{llll}
\rho_i(t)\hspace{3.6cm} && \mbox{Probability $(1 - 3/N)$} \nonumber \\
	   \end{array}
	   \right. \\
&& \left\{ \begin{array}{ll}
\rho_i(t) &\hspace{0.3cm} \mbox{Probability $(1-p^R_{i-1})/N$} \nonumber \\
\rho_i(t)+ (1-\rho_i(t))\rho_{i-1}(t) &\hspace{0.3cm}
 \mbox{Probability $p^R_{i-1}$/N } \\
	   \end{array}
	   \right. \\
\rho_i(t+1)~~ = ~~&& \left\{ \begin{array}{llll}
\rho_i(t)\hspace{2.2cm} &&& \mbox{Probability $(1-p^R_i - p^L_i)/N$}
\nonumber \\
\rho_i(t)\rho_{i-1}(t)\hspace{2.2cm} &&& \mbox{Probability $p^L_i$/N}
\nonumber \\
\rho_i(t)\rho_{i+1}(t)\hspace{2.2cm} &&& \mbox{Probability $p^R_i$/N}
\nonumber \\
	   \end{array}
	   \right. \\
&& \left\{ \begin{array}{lll}
\rho_i(t) && \mbox{Probability $(1-p^L_{i+1})/N$} \\
\rho_i(t) + (1-\rho_i(t))\rho_{i+1}(t)&& \mbox{Probability $p^L_{i+1}/N$} \\
	   \end{array}
	   \right. 
\label{eq:hops1}
\end{eqnarray}
The first term in Eq. (\ref{eq:hops1}) represents
the probability that a site other than $i$ is picked
at time step $t$, whereas subsequent terms represent
probabilities that the density variable at site $i$
is updated as a consequence of site $i-1,i$ or $i+1$
being picked. An overall factor of $1/2$ in these
transition probabilities (arising from the fact that
the choice to update a site or potential is made
with probability $1/2$ at every time step), has been
absorbed into a rescaling of time.

The hopping rates
$p^R_{i-1},~p^R_{i},~p^L_{i},~p^L_{i+1}$ are all
explicitly functions of time and are determined by
the instantaneous state of the potential. This state
is determined by a corresponding equation for the
dynamics of the potential field or equivalently of
the stochastic variable $\eta(t)$ which appears in
the definition of $V(x,t)$.  We must specify two
averages in the steady state.  One is the average
over thermal noise given a {\em particular} stochastic
history of potential flips. The other is the average
over all stochastic histories.  We are interested in
those attributes of the system which characterize its
steady state.  The dynamics of potential and particle
are partially decoupled. The potential state influences
the particle hopping rate but the potential flips
independently of the particle state.
\begin{figure}
\includegraphics[width=\columnwidth]{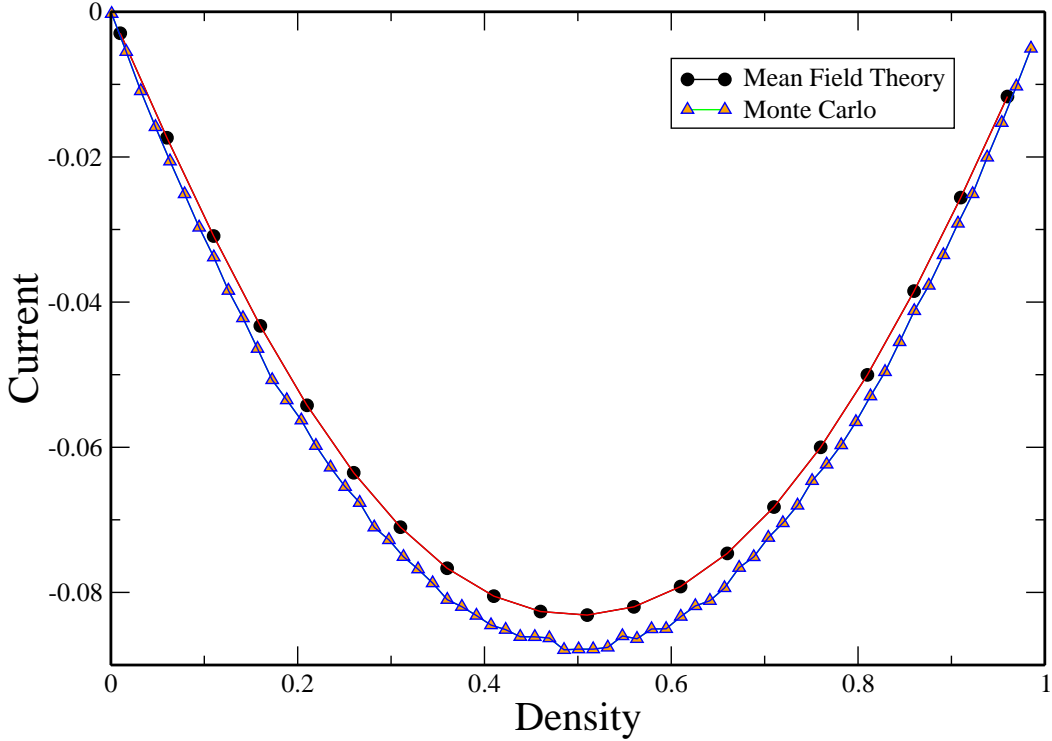}
\caption{The current-density relation as obtained through Monte
Carlo simulations and the mean field theory described in
the text, assuming $r=0.05$, $P_{01} = 0.03$ and
$P_{10} = 0.04$.
\label{fig:mft}
}
\end{figure}
The mean-field approximation made
is the following: Whenever a product such as
$\rho_i(t)\rho_{i+1}(t)$ is to be averaged
over either thermal noise or potential flip
histories (with the appropriate averaging
operation denoted by $\langle\cdot\rangle$), we
replace $\langle\rho_i(t)\rho_{i+1}(t)\rangle$ by
$\langle\rho_i(t)\rangle\langle\rho_{i+1}(t)\rangle$.
This approximation truncates the hierarchy of coupled
correlation functions by representing correlation
functions of all higher-order products of density
fields in terms of single-site averages.
Our mean-field theory is formulated for the following
limit: If the potential fluctuates over a microscopic
time-scale which is much faster than characteristic
diffusion time scales over a single period, it is
legitimate to {\em average} the rates.  The discrete
equation Eq. (\ref{eq:hops1}), can then be converted
into a first-order non-linear system of differential
equations in time.  These equations are
\begin{equation}
\label{eq:continuum}
\frac{d\rho_i(t)}{dt} = \tau^1_i\rho_i(t) + \tau^2_i\rho_{i-1}(t) +
\tau^3_i\rho_{i+1}(t) + \tau^4_i\rho_{i-1}(t)\rho_{i}(t)
+ \tau^5_i\rho_{i}(t)\rho_{i+1}(t)
\end{equation}
where $i$ runs over the sites in a single period. Given $P_{01}$ and
$P_{10}$, the definitions of $\tau^1_i \ldots \tau^5_i$ are the 
following:
$\tau^1_i = - {\bar p}^L_i- {\bar p}^R_i, 
\tau^2_i = {\bar p}^R_{i-1},
\tau^3_i = {\bar p}^L_{i+1},
\tau^4_i = {\bar p}^L_{i} + {\bar p}^R_{i-1}$ and
$\tau^5_i = {\bar p}^R_{i} + {\bar p}^L_{i+1}$.
The bars denote an average over the rates.
The set of equations Eq.(\ref{eq:continuum}) represent the mean-field
treatment of the case in which the average over stochastic histories of
potential flips has been performed before the average over thermal
histories. 
\begin{figure}
\includegraphics[width=\columnwidth]{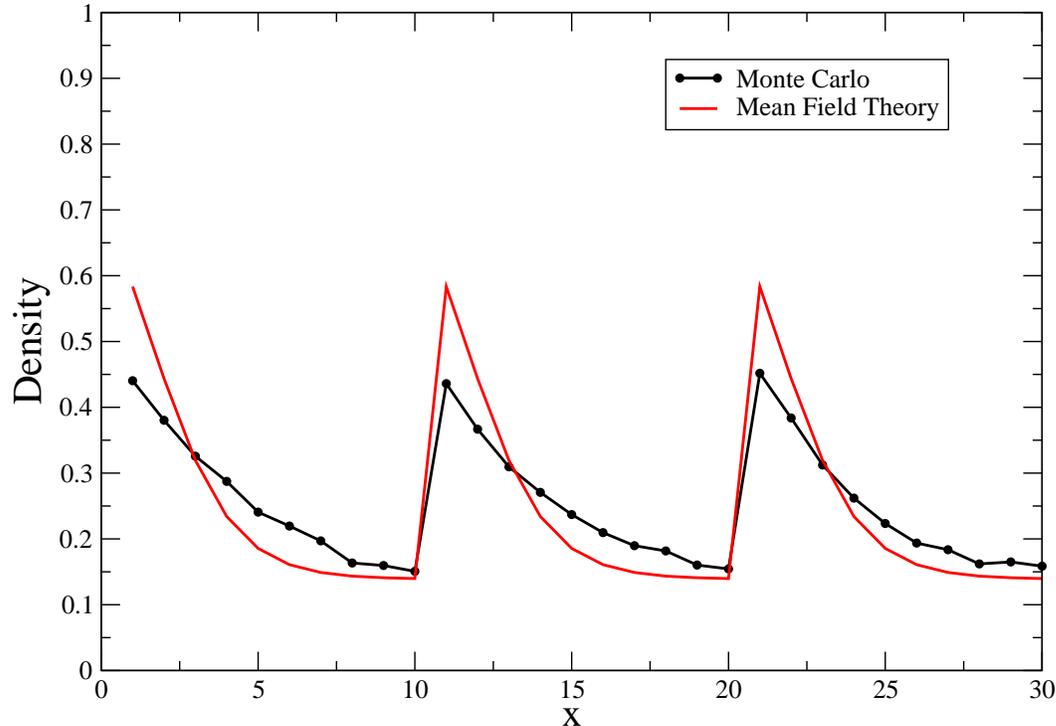}
\caption{The comparison between local densities in
each period as obtained by averaging the Monte
Carlo data and predictions of the mean field theory 
described in the text, assuming $r=0.05$
$P_{01} = 0.03$ and $P_{10} = 0.04$. The mean density is 0.25.
\label{fig:mft1}
}
\end{figure}
The steady state in mean-field theory is obtained by setting the time
derivatives $d\rho_i(t)/dt$ to zero. The resulting equations 
are to be solved for the W sites within a period,
Given a mean field solution for the densities, the time-averaged
current in the mean-field approximation can be obtained.
\begin{figure}
\includegraphics[width=\columnwidth]{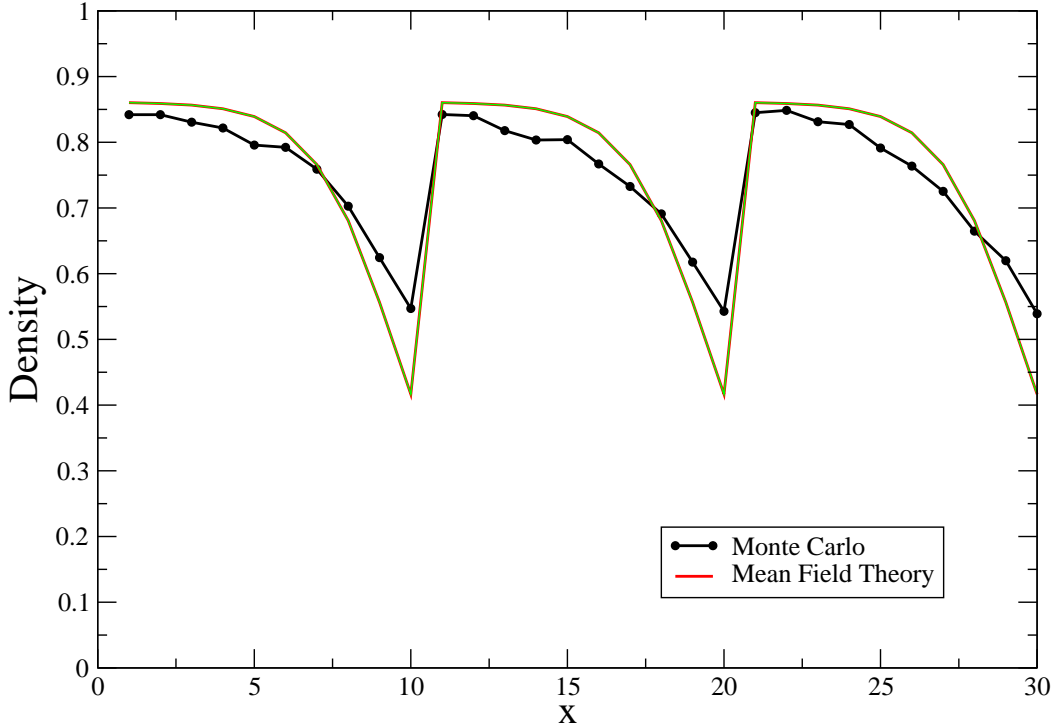}
\caption{The comparison between local densities in
each period as obtained by averaging the Monte
Carlo data and predictions of the mean field theory 
described in the text, assuming $r=0.05$
$P_{01} = 0.03$ and $P_{10} = 0.04$. The mean density is 0.75.
\label{fig:mft2}
}
\end{figure}
Fig \ref{fig:mft} illustrates the comparison of the
fundamental diagram of the system as obtained through
the mean field theory with the Monte Carlo data.
The density profiles for densities 0.25 and 0.75 are
shown in Figs. \ref{fig:mft1}  and \ref{fig:mft2}
together with the results of a direct numerical
simulation. It can be seen that the mean-field
theory in the averaged-rate limit used above predicts
the currents and the density profiles to reasonable
accuracy. At the level of the currents, the agreement
between the mean-field theory and the simulation
results is certainly passable. The density profiles 
appear qualitatively accurate but are incorrectly 
rendered in quantitative terms.

\section{Boundary-Driven Phase Transitions}
\label{sec:bdpt}
The periodic boundary condition assumed in previous
sections is clearly incorrect in the biological
context, where motors are loaded and unloaded along
the cytoskeleton at rates dictated by the local
chemistry. As the two ends of a cytoskeletal filament
loaded with motors carrying cargo are potentially
well separated in space, their chemical
environments need not be identical. It is thus
possible that motor {\em loading} and
{\em unloading} rates could be different at either
end of the filament.  The generalization of
ratchet models for the motion of interacting motors,
extended to allow for open boundary conditions,
could potentially allow for the boundary injection
and removal of motors. This injection and removal will
compete, in general, with the bulk equilibration via
Langmuir kinetics\cite{andreaprl,andreapre,hinsch05}
of the motor density along the filament, but we
will ignore the possibility that motors are lost or
gained along the filament, accounting only for their
entry and exit at the boundaries.  For recent work
which incorporates both bulk non-conservation and the
effects of geometry, see Refs.
\cite{lipprl01,nieepl02,niepre04,klu01,klu05}.

For simulations with open boundary conditions, we
attach two boundary sites to the open chain with $N$
sites; the total number of sites is then $N+2$. These
boundary sites are filled with probability $\alpha$ (at
the right boundary) or $\beta$(at the left boundary). 
Hopping rules at these special sites are chosen to
be consistent with the direction of the net current
in the absence of boundary driving and motors at these
singled-out sites always hop unidirectionally.
\begin{figure}
\includegraphics[width=\columnwidth]{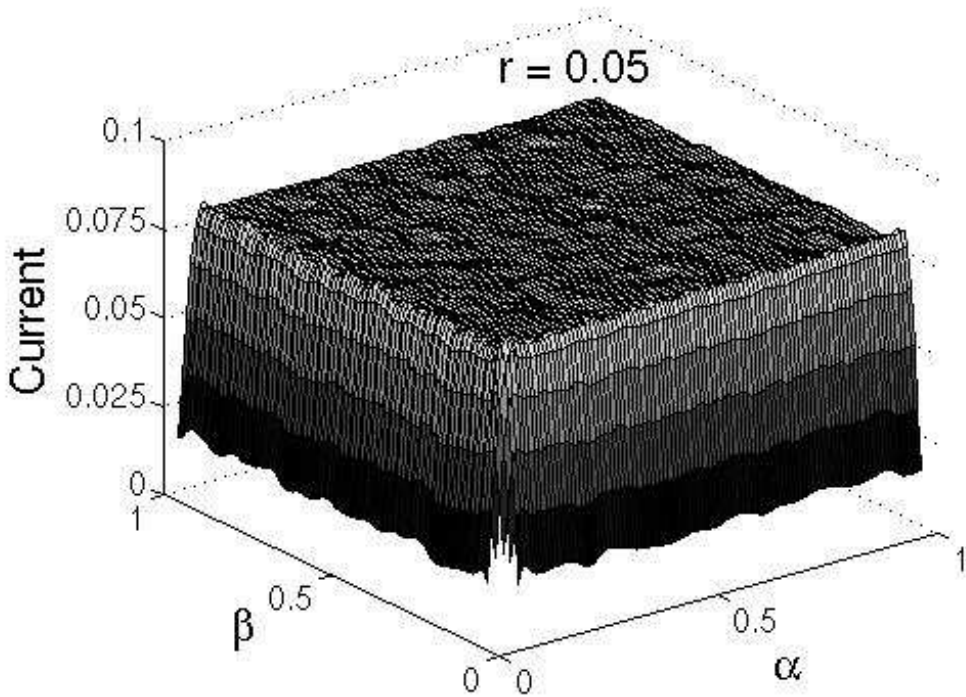}
\caption{Current versus the boundary and injection rates
$\alpha$ and $\beta$, obtained through Monte Carlo
simulations for parameter values $N = Lw = 128*10$ sites,
$r=0.05$, $P_{10} = 0.04$ and $P_{01} = 0.05$. 
\label{fig:jdriv}
}
\end{figure}
\begin{figure}
\includegraphics[width=\columnwidth]{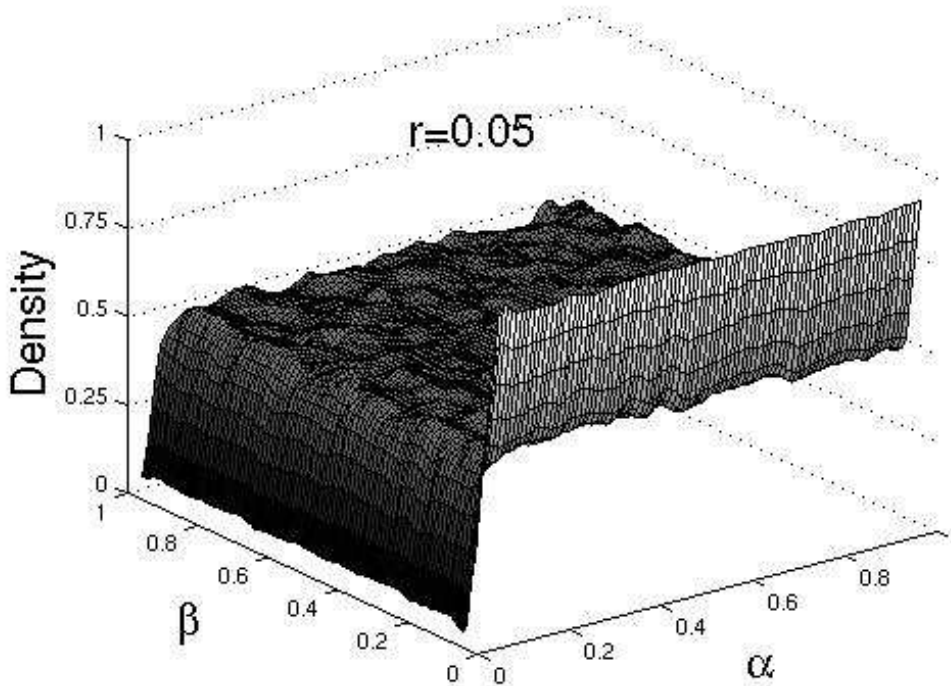}
\caption{Current versus the boundary and injection rates
$\alpha$ and $\beta$, obtained through Monte Carlo
simulations for parameter values $N = Lw = 128*10$ sites,
$r=0.05$, $P_{10} = 0.04$ and $P_{01} = 0.05$. 
\label{fig:dendriv}
}
\end{figure}
Results for this model are shown in Fig \ref{fig:jdriv}
which exhibits the current in the system $J$ as a function of
the input and output rates, $\alpha$ and $\beta$. Note
the striking feature of the plot --  the remarkable
independence of $J$ on $\alpha$ and $\beta$ for a wide
range of these parameters.
For this range of input and output rates, the current
is not only {\em independent} of the rates at which
motors are added or subtracted at both ends but the
current being passed through the system is pegged
at its {\em maximum} value.  Fig. \ref{fig:dendriv}  shows
the steady state densities at varying
input and output rates. 

Note the existence of three distinct phases: (I) a
phase in which the current and density can be varied by
changing only $\beta$ and is independent of $\alpha$,
(II) a phase in which the current and density can be
varied by changing only $\alpha$ and is independent
of $\beta$, (III) a ``constant current'' phase
in which the current attains its maximum possible
value and is independent of both $\alpha$ and $\beta$.
The boundary driven ASEP has a phase diagram which is
qualitatively very closely similar, with a low density
phase, a high density phase and a intermediate constant
current phase\cite{robin}. It should, in principle
be possible to investigate the same effects described
in Refs. \cite{andreaprl,andreapre,hinsch05}, where
Langmuir kinetics competes with one-dimensional transport
to yield a phase diagram with complex structure.
We do not, however, address this interesting problem here.

\section{Motor Microtubule Pattern Formation}
The mitotic spindle of a dividing eukaryotic cell is
a self-organized  structure at the
sub-cellular scale. Such structures are usefully
thought of as patterns, by which we mean spatially
inhomogeneous yet stable steady states, defined
through the interaction of motors and microtubules.
Experiments on centrosome-free fragments of the
cytosol containing both motors and microtubules
obtain self-organized radial structures called
asters.  Single asters,
in addition to other complex patterns such as
vortices, disordered aster-vortex mixtures and
lattices of asters and vortices, are also seen {\it
in vitro}, in experiments on mixtures of molecular
motors and microtubules in a quasi-two-dimensional
geometry\cite{nedelecsurreymaggsleibler}.

What physical processes stabilize such structures?
Direct molecular dynamics simulations which incorporate
any level of molecular-scale realism are currently
incapable of tacking the pattern formation problem. One
must then rely on approximate models for such systems,
again keeping in mind the necessity of maintaining
contact with biological reality: our models must be
as simple as possible but no simpler. This problem
has been studied extensively over the past 5 years
or so with significant contributions from several
groups \cite{leekardar,leekim,tannie,kruse3,aranson}.
This section summarizes work in this direction first
presented in Ref. \cite{sumithra}.

In this section, hydrodynamic equations of motion
for a coarse-grained field representing the local
orientation of microtubules as well as for local motor
density fields are described.  Our model treats motors
attached to microtubules differently from motors which
diffuse freely in solution.  Motors which move on
microtubules are referred to as  ``bound''  motors,
while those which diffuse in the ambient solvent are
referred to as ``free'' motors. These are described
by coarse-grained fields denoted by $m_b$ and $m_f$
respectively and obey different  equations of motion.

We will take microtubules to be oriented by complexes
of bound motors, yielding patterns at large scales.
Our results are: We obtain a single vortex as a stable
final state for large motor densities in some regimes.
However, in other regimes, asters are favoured.
A ``lattice of asters'' state is stabililized in
our model through a low-order relevant term in the
equation of motion for the microtubule orientation.
On small systems, constraints due to confinement
favour a small number of asters, whose number can be
increased systematically as parameters are varied.
We have also calculated the distribution of free
and bound motors in asters and vortices obtained in
our model; we derive an exponential decay of bound
motor densities away from aster cores, modulated by
a power-law in which the exponent of the power law
depends in a non-universal way on dynamical parameters.

In the absence of interconversion terms changing
a bound motor to a free motor, $m_b$  obeys a
continuity equation involving the  current of motors
transported along the microtubules. The free motor
field $m_f$ obeys a diffusion equation with a diffusion
constant $D$.  These two fields are  coupled  through
mechanisms which convert ``free''  motors to ``bound''
motors and vice versa. We will take $\ftob^{\prime}$
and $\btof^{\prime}$ to be the rates at which free
motors become bound motors (``on'' rate) and vice-versa
(`off'' rate).

Our aim is to write down an minimal set of equations
capable of both describing the variety of patterns
formed in such interacting motor-microtubule mixtures.
We are guided both by symmetry, as is appropriate for a
fully non-equilibrium system in which detailed balance
based on rates derivable from a hamiltonian does not
exist, as well as by considerations of simplicity: of
an infinity of possible non-equilibrium terms allowed
in our equations of motion, we choose the simplest.
In appropriately scaled units, the equations then
are 
\be
\label{scalefden}
\partial_t m_f = \nabla^2
m_f - \ftob m_f + \btof m_b \ee 
\be
\label{scalebden}
\partial_t m_b = -\nabla \cdot (m_b {\bf T}) + \ftob
m_f - \btof m_b \ee \be \label{scalet} \partial_t
{\bf T} = C {\bf T} (1 - T^2) + m_b \nabla^2 {\bf T}
+ \epsilon \nabla m_b \cdot \nabla {\bf T} + \kappa
\nabla^2 {\bf T}+S \nabla m_b 
\ee

Free and bound motor density profiles in vortex
and aster configurations can be obtained from the
above equations.  We set the time derivatives to zero
{\it i.e.}\/ $\partial_t m_f = \partial_t m_b = 0$,
obtaining \be \label{eq1} \nabla^2 m_f - \nabla \cdot
(m_b {\bf T}) = 0, \ee \be \label{eq2} \nabla^2 m_f +
(\btof m_b - \ftob m_f) = 0.  \ee 
For a single vortex i.e. ${\bf T}$ = $\hat \theta$,
we may assume radial symmetry and thus $m_f = m_f(r)\
\textnormal{and}\ m_b = m_b(r)$.  
We then obtain,
\bea m_f(r) = c_1 + c_2 \ln(r), \eea where $c_1$
and $c_2$ are constants to be determined by boundary
conditions and normalization.
The relation $ \btof m_b - \ftob m_f = 0  $
yields
\be m_b(r) = \frac{\ftob}{\btof}
(c_1 + c_2 \ln(r)).
\ee

For the motor distribution about a single aster
i.e. ${\bf T} = -{\bf \hat r}$, we again assume radial
symmetry for the bound and free motor densities.
The boundary condition that the total motor current
vanishes at the boundary implies
\be
\label{astercondition}
\partial_r m_f(r)  = - m_b(r).  
\ee 
Thus we have
\be 
\label{asterfree} \partial_r^2 m_f +
(\frac{1}{r} - \btof) \partial_r m_f - \ftob m_f = 0.
\ee 
The general solution to the equation above is
a combination of confluent hypergeometric functions and
can be written in terms of the two solutions of the
hypergeometric Kummer equation, modulated by an
exponential.  It is useful
to define a quantity $p$ given by 
\be p = \frac{1}{2}(1
- \frac{\btof}{\sqrt{\btof^2+4\ftob}}).  
\ee
Note that
$0 \le p \le 0.5 $ with $\btof,\ftob \ge 0$. 

The asymptotics is obtained using an integral 
representation for the appropriate hypergeometric functions.
Our final result for the motor distribution about a fixed
aster configuration is
\bea
m_f(r)  &\sim& c_1 \frac{e^{-r/\xi}}{{(\btof^2+4\ftob)}^{p/2} r^p} \nn \\
m_b(r) &\sim& c_1 \frac{e^{-r/\xi}}{{(\btof^2+4\ftob)}^{p/2} r^p}
(\frac{p}{r}+\frac{1}{\xi}), \nn
\eea
with $\xi^{-1} = \Big|\frac{(\btof - \sqrt{(\btof^2+4\ftob})}{2}\Big| = 
\Big|\frac{p\btof}{2p-1}\Big|$.
The correlation length $\xi$ and the power-law exponent $p$
depend on $\ftob$ and $\btof$.
We see that the bound motor density in the aster case has an
exponential fall modulated by a power-law tail.

We also solve Equations\
\ref{scalefden},\ref{scalebden} and \ref{scalet}
numerically on an $L \times L$ square grid indexed by
$(i,j)$ with $i = 1,\ldots L$ and $j = 1,\ldots L$.
The equations for the free and bound motor densities
are evolved using an Euler scheme.  We impose the
boundary condition that no current (either of free or
bound motors) flows into or out of the system. This
condition is easily imposed by setting the appropriate
current to zero.  The {\bf T} equation is differenced
through the Alternate Direction Implicit (ADI) operator
splitting method in the  Crank-Nicholson scheme.
\begin{figure}
\includegraphics[width=\columnwidth]{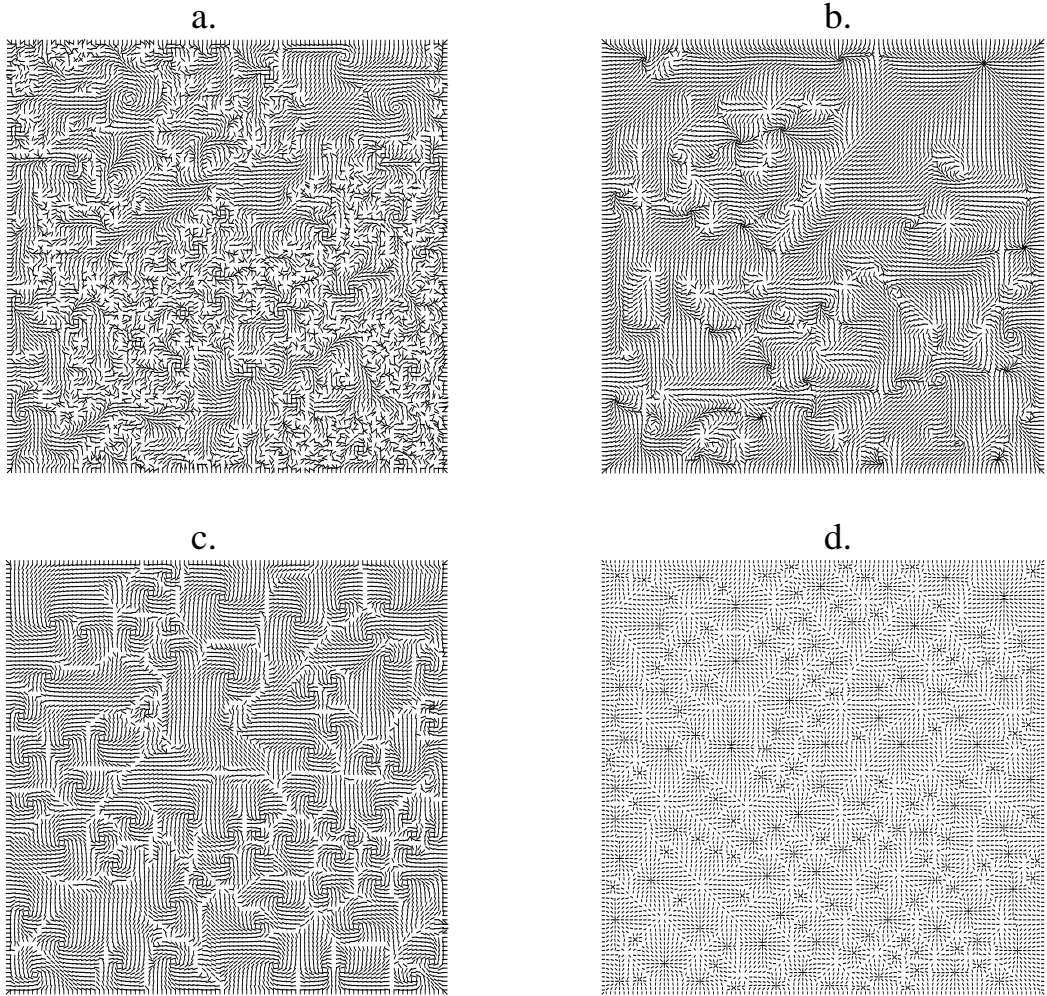}
\caption{
Steady state configurations in our model 
at different parameter values (see text):
(a) Disordered states
obtained at very low motor densities [$m = 0.005$
$\epsilon = 0.5$ and $S = 0$];
(b) Aster-vortex mixture obtained at
[$m = 0.01$, $\epsilon = 0.5$ and $S = 0$];
(c) Lattice of vortices at
[$m = 0.05$, $\epsilon = 5$ and $S=0.001$];
(d) Lattice of asters
obtained at [$m = 0.5$,
$\epsilon = 1$ and $S=1$]
\label{fig:hund}
}
\end{figure}

Our simulations are on lattices of several sizes,
ranging from $L=30$ to $L=200$. We vary the motor
density in the range 0.01 to 5 in appropriate
dimensionless units.  We work with two different types
of boundary conditions on the $T$ field. In the first,
which we refer to as reflecting boundary conditions,
the microtubule configuration at the boundary sites is
fixed to point along the inward normal.  In the second,
which we refer to as parallel boundary conditions,
microtubule orientations at the boundary are taken
to be tangential to the boundary.  In both these sets
of boundary conditions, the state of the boundary $T$
vectors is fixed and does not evolve. The total number
of motors, initially divided equally between free and
bound states and distributed randomly among the sites,
is explicitly conserved.
\begin{figure}
\includegraphics[width=\columnwidth]{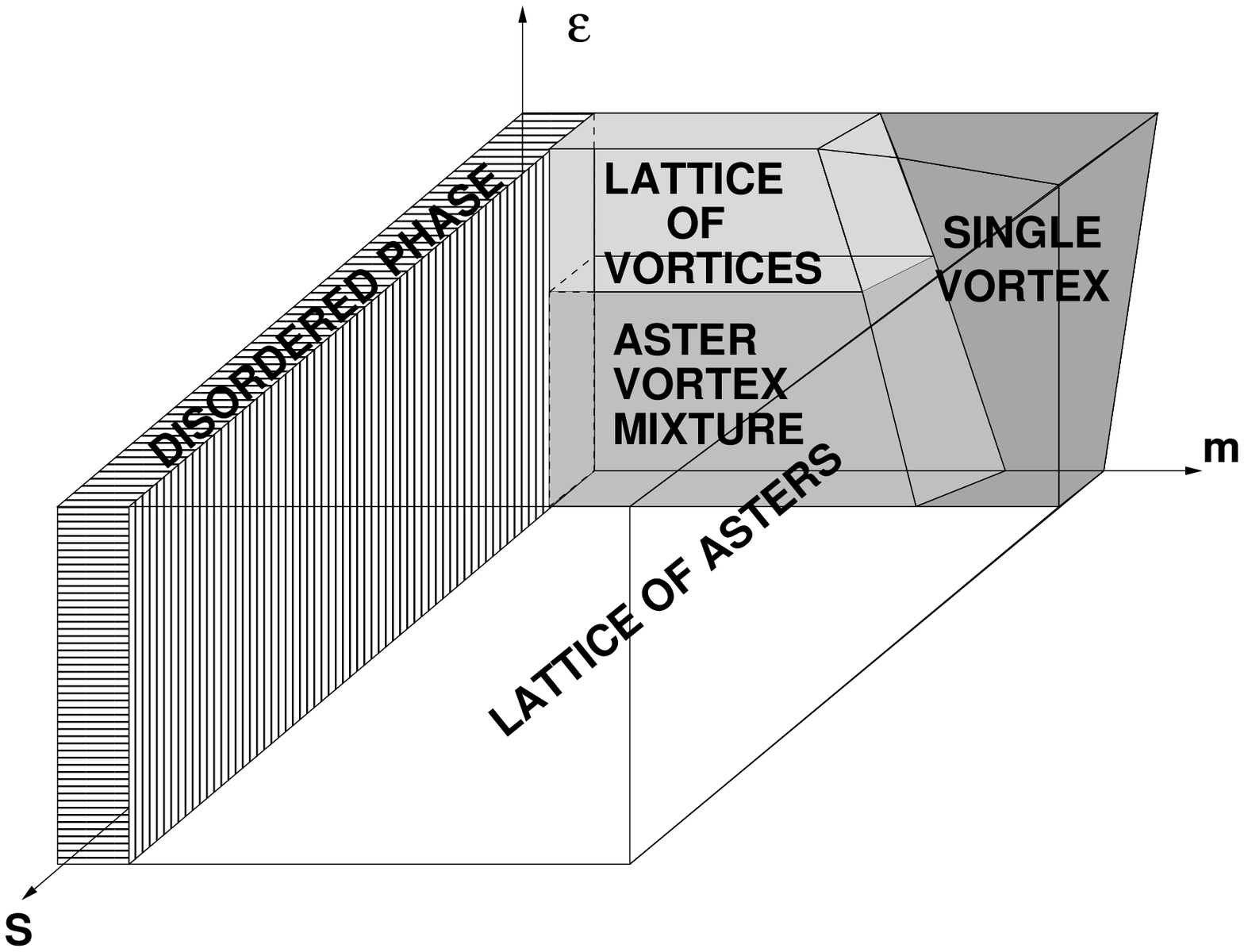}
\caption{
Qualitative map of steady states illustrating how
different states, the disordered state, the
aster-vortex mixture state, the lattice of
vortices state, the single vortex, the lattice of asters,
dominate in different regimes of parameter space;
for a definition of parameters see text. The
parameter $\epsilon$ is plotted on the $y$ axis,
with the total motor density $m$ plotted on the
$x$ axis. The parameter $S$ extends out of the
$\epsilon-m$ plane. Of the states shown, the lattice
of asters is obtained generically for non-zero $S$ (out of the
plane of the figure), whereas the other states are associated
with the $S=0$ plane, although they appear to survive provided
$S$ is small enough.}
\label{fig:phase}
\end{figure}

Figures \ref{fig:hund}(a) -- (d) depict four
stable configurations obtained in different
regimes of parameter space for an $L=100$ lattice.
Fig. \ref{fig:hund}(a) shows a disordered arrangement
of microtubules obtained at very low motor densities
($m = 0.005$) with $\epsilon = 0.5$ and $S = 0$.
Figure \ref{fig:hund}(b) shows an aster-vortex
mixture obtained at $m = 0.01$ at the same values of
$\epsilon$ and $S$.  This figure is to be contrasted
to Fig. \ref{fig:hund}(c), obtained at $m = 0.05$,
taking $\epsilon = 5$ and $S=0.001$. Note the absence
of asters in this regime of parameter space. Finally,
Fig \ref{fig:hund}(d), obtained with $m = 0.5$,
$\epsilon = 1$ and $S=1$, illustrates a lattice of
asters, with asters being the only stable defects
present. We can vary the sizes and numbers of asters
obtained in configurations such as the one shown in
Fig. \ref{fig:hund}(d), by changing $S$. A larger
$S$ yields a large number of small asters, while
smaller values of $S$ yield a smaller number of large
asters\cite{nedelecsurreymaggs}.

Our results are summarized in Fig. \ref{fig:phase}
which shows the states which dominate in the
three-dimensional space spanned by $\epsilon$, $S$
and $m$.  For $S=0$, we obtain disordered/aster-vortex
mixture states at low motor density, which become
a lattice of vortices at somewhat higher motor
densities.  Large values of $\epsilon$ ($\epsilon
\geq 1$) yield well-formed vortex-like configurations
while small $\epsilon$ yields structures better
described as aster-vortex mixtures.  At intermediate
values of $\epsilon$ and $m$, spirals rather than
vortices appear to dominate.  At large $m$, with
$S=0$ and large $\epsilon$, a single vortex is
obtained\cite{leekardar,leekim}.

For non-zero but small $S$, these states appear
to continue out of the $S=0$ plane but are rapidly
replaced by a lattice of asters for larger $S$. A cut
of Fig. \ref{fig:phase} at finite $S$ yields disordered states at
small $m$ and a lattice of asters at larger $m$. We
can thus understand the sequence of patterns formed
upon increasing $m$ in mixtures of kinesin constructs
with microtubules in terms of a trajectory which
begins in the $S=0$ (or $S$ sufficiently small)
plane in the disordered phase and transits between the
aster-vortex mixture and the lattice of vortices (both
of which lie in this plane) as $m$ is increased. As
$m$ increases further and the effects of the $S$ term
become important, such a trajectory moves out towards
non-zero $S$, encountering the lattice of asters.

We have also examined the effects of changing the
motor processivity, a quantity proportional to the
ratio of $\ftob$ to $\btof$. Smaller values of this
ratio are appropriate to molecular motors such as NCD.
At $\ftob = 0.005, \btof = 0.05$, we find that the
disordered regime shown in Fig. 1 expands, so that at
equivalent values of $m$ disordered states occupy much
of the domain associated previously with the lattice
of vortices.  Whereas kinesins follow the sequence
{\em disordered -- lattice of vortices -- aster vortex
mixture -- lattice of asters} as the $m$ is increased,
a mixture of microtubules with NCD motors bypasses the
lattice of vortices altogether, transiting directly
from the disordered state to the lattice of asters in
the experiments\cite{surreynedelecleiblerkarsenti}.
In terms of Fig. \ref{fig:phase}, the expanded regime
of disordered states for the NCD motor suggests
that patterns such as the lattice of vortices and
the aster-vortex mixture may be inaccessible at the
motor densities at which the experiments are done,
since the effects of the $S$ term might be expected
to dominate at large $m$.

It is interesting to note that the generic state
which is obtained at large $m$ is a lattice of asters
for non-zero $S$, in contrast to the predictions of
the earlier model of Lee and Kardar, which indicates
that the large motor density state is always a single
vortex.  This feature, a direct consequence of the
presence of the crucial $S\nabla m$ term, agrees
with experiment. Further results, including a discussion
of motor-motor interactions in the context of motor-
microtubule pattern formation, qualitative ``free-energy''
based arguments for the stability of patterns and 
a detailed discussion of the effects of confinement can be
found in Ref. \cite{sumithra}.

\section{Conclusions}

The interacting thermal ratchet model discussed in
the first part of this paper introduces a primitive
level of biological realism into models for the
motion of interacting motor proteins.  While the model
itself may not appear appreciably more realistic
than the ASEP itself, the use of ratchet models
generalized to include interactions is attractive,
since such models incorporate the {\em true}
reason for the symmetry breaking which occurs
when molecular motors move unidirectionally in a
Brownian environment. This reasoning is obscured
in the currently popular ASEP-based models, where
the symmetry breaking which leads to a non-trivial
current is imposed by hand, through the definition of
the hopping rates.  Making the connection to ratchet
models also facilitates the understanding of many
issues relating to the coherence or reliability of
transport\cite{jayan1} in motor systems as well as of
efficiency in energy transduction\cite{efficiency}.
The exclusion processes simply lack the necessary
structure for such discussions to be meaningful.

We have also described calculations which relate to
a non-equilibrium pattern formation problem: the
formation of patterns in mixtures of microtubules
and molecular motor constructs. The model is able
to reproduce virtually all the patterns obtained in
the experiments, but has the advantage over direct
simulations that the number of parameters which need
to be included is small. We have also been able to
derive many of the features of the patterns which
form, including the sequence of patterns which are
obtained as the motor density is increased. Our model
rationalizes several features of the experiments,
in many cases for the first time. These include:
(i) the sequence of patterns obtained as the motor
density is increased, (ii) the prevalence of the
lattice of asters state and (iii) the difference in the
sequence of patterns formed by conventional kinesins
and the NCD motor.  Many further features of these
equations are currently being explored.

One final point relates to the nature of steady
states which are obtained in the ratchet models
generalized to include interactions between motors
and has to do with the possibility of ``robustness''
in these models. Barkai and Leibler\cite{barkai} and
Alon {\it et.  al.}\cite{alon} study the chemotaxis
network of {\it E.  Coli}, suggesting that this network
exhibits {\em exact} and {\em robust} adaptation,
over a wide range of variation of parameters. The
robustness of adaptation in this case is an example
of biochemical robustness, since it has its origins in
specific features of the biochemical network underlying
chemotaxis in {\it E.  Coli}.  The ``tensegrity'' of
the cytosketal network of living cells can be thought
of as another form of robustness.  Such robustness
of the structural elements in the cell to mechanical
perturbations is distinct from biochemical robustness.
It is thus interesting to ask the following question:
Are other manifestations of robustness possible and
are there biological situations in which they may
be relevant?

We suggest another intriguing possibility for robust
behaviour in biological systems, the {\em robustness
of certain non-equilibrium steady states of biological
systems to wide classes of physical perturbations}.
A simple illustrative example in the context of the
boundary driven interacting ratchet model is the
independence of the steady state current on $\alpha$
and $\beta$ for a very large range of such parameter
values. This indicates that the current is insensitive
to fairly large fluctuations in the input and output
rates and is, moreoever, pegged to the largest possible
value it can take. The robustness here is for {\em
physical} reasons, essentially having to do with the
one-dimensional character of this steady state and the
fact that the system can adjust its density profile
to accomodate changes in the boundary driving rates.
While allowing for Langmuir kinetics in the bulk
in this specific case will generically destroy such
behaviour in the thermodynamic limit, small systems
should exhibit the same qualitative behaviour.  It
remains to be seen if such ``steady-state robustness''
or ``physical robustness'' is in fact a feature of some
aspects of {\it in vivo} cellular function, independent
of the models we contrive to describe such function. It
would be interesting to look at other examples of
non-equilibrium steady states in biological systems,
possibly those which involve cell-scale flows (as,
for example, in cell streaming), to see if this idea
might find support.

\section{Acknowledgements} 
The work described in the first section was done in
collaboration with Yashar Aghababaie and Michael
Plishke at Simon Fraser University, with research
supported by the NSERC of Canada. The work described
in the third part, relating to the motor-microtubule
pattern formation problem was done in collaboration
with Sumithra Sankararaman at the Institute of
Mathematical Sciences and P.B. Sunil Kumar at the IIT,
Madras and was partially supported by the DST (India).
I am grateful to these collaborators for all I have
learnt from them. At various stages, I have benefited
from conversations with Michael Wortis, Jacques Prost, 
Mustansir Barma, Deepak Dhar, Arun Jayannavar,
Madan Rao, Sriram Ramaswamy and Amit Kumar Bhattacharjee.
I would also like to thank Debashish Chowdhury for
organizing a very stimulating and topical conference
on these and related issues where this work was presented.


\end{document}